\documentclass[floatfix,reprint,showpacs,preprintnumbers,amsmath,amssymb,aps,prl,superscriptaddress,longbibliography,showkeys,nobibnotes]{revtex4-2}

\usepackage[english]{babel}

\usepackage{physics}
\usepackage{graphicx}
\usepackage{placeins}
\usepackage[colorlinks=true, allcolors=blue]{hyperref}
\usepackage{soul}
\usepackage[utf8]{inputenc}

\newcommand{\anh}{\ensuremath{\hat{a}^{\vphantom{\dagger}}}_\vec{p}}
\newcommand{\cre}{\ensuremath{\hat{a}^\dagger}_\vec{p}}
\renewcommand{\vec}{\mathbf}
\newcommand{\beqar}{\begin{eqnarray}}
\newcommand{\eeqar}{\end{eqnarray}}

\begin{document}
\title{Ab initio path-integral Monte Carlo results for the one-particle spectral function of the warm dense electron gas}
\author{Paul Hamann}
\email{p.hamann@hzdr.de}
\affiliation{Institute of Radiation Physics, Helmholtz-Zentrum Dresden-Rossendorf (HZDR), D-01328 Dresden, Germany}
\affiliation{Institut f\"ur Physik, Universit\"at Rostock, D-18057 Rostock, Germany}

\author{Michael Bonitz}
\affiliation{Institut f\"ur Theoretische Physik und Astrophysik, Christian-Albrechts-Universit\"at zu Kiel, D-24098 Kiel, Germany}

\author{Jan Vorberger}
\affiliation{Institute of Radiation Physics, Helmholtz-Zentrum Dresden-Rossendorf (HZDR), D-01328 Dresden, Germany}

\author{Tobias Dornheim}
\affiliation{Institute of Radiation Physics, Helmholtz-Zentrum Dresden-Rossendorf (HZDR), D-01328 Dresden, Germany}
\affiliation{Center for Advanced Systems Understanding (CASUS), D-02826 G\"orlitz, Germany}
\date{\today}

\begin{abstract}
We present quasi-exact \emph{ab initio} path-integral Monte Carlo results for the Matsubara Green's function of the uniform electron gas (UEG) at finite temperature over a broad range of coupling strengths ($r_s=1,\dots,10)$. We further extract the static self-energy $\Sigma_\infty(p)$ and perform an analytic continuation for spectral function $A(p,\omega)$, conclusively ruling out the possibility of distinct satellite features at these conditions. In addition, our work opens up intriguing avenues to study the single-particle spectrum and density of states of real warm dense matter systems based on first principles.
\end{abstract}
\maketitle

The archetypical uniform electron gas (UEG) is one of the most important model systems in physics and quantum chemistry~\cite{quantum_theory,review}. While the UEG has been a central subject in the development of many-body perturbation theory \cite{bohm53,gellmannbrueckner,hedin}, an accurate description at arbitrary coupling strength only became possible with the advent of Quantum Monte Carlo (QMC) methods \cite{ceperley80}.

In the last decade, there has been increased interest in studying the properties of warm dense matter (WDM) \cite{Graziani2014,vorberger2025roadmapwarmdensematter}, a loosely defined term for the extreme conditions of simultaneously high density ($r_s \sim a_B$, with $r_s$ being the Wigner-Seitz radius) and temperature ($k_B T \sim E_F$, with $E_F$ being the Fermi energy) that are expected in the interior of celestial objects such as giant planets as well as white and brown dwarfs \cite{Vorberger2007,Nettelmann2016,Militzer2008,Soubiran2017,Chabrier2000}. In addition, WDM is also becoming increasingly relevant for cutting-edge technological applications such as inertial confinement fusion (ICF) experiments~\cite{hu_ICF,Hurricane_RevModPhys_2023}.
As a consequence, extreme states of matter are routinely realized in the laboratory using a variety of techniques~\cite{Falk2018,vorberger2025roadmapwarmdensematter} such as laser compression \cite{Pascarelli2023}. The accurate theoretical description of WDM requires a full quantum-mechanical treatment while taking both electronic correlations and the finite temperature into account, which is highly challenging~\cite{Graziani2014,new_POP}.

Recent advances in the development of path-integral Monte Carlo (PIMC) methods~\cite{groth2017_prl,Dornheim2017,review} have allowed for simulations of the UEG (and very recently also real warm dense matter systems such as hydrogen~\cite{Filinov_PRE_2023,Dornheim_MRE_2024} and beryllium~\cite{Dornheim_NatComm_2025}) over a broad range of WDM parameters without relying on the previously used fixed-node approximation~\cite{Brown_PRL_2013}.
However, being formulated in the imaginary-time domain, PIMC only gives direct access to static quantities in equilibrium. Spectral quantities, which typically carry richer information and are directly related to experimental measurements, can only be accessed in terms of imaginary-time correlation functions. This requires an additional analytic continuation to the real frequency domain \cite{Gubernatis1991,Sandvik1998,Vitali2010,chuna2025noiselesslimitimprovedpriorlimit}. Overcoming this well-known but ill-defined problem for the density-density correlation function has recently led to the first ab-initio PIMC results for the dynamic structure factor and various related quantities \cite{Dornheim2018_prl,Groth2019,Hamann2020,Chuna_PRB_2025,Filinov_PRB_2023}.

In addition to such two-particle correlation functions describing scattering processes, PIMC simulations involving a single trajectory with open ends (the eponymous ``worm'' of Boninsegni \textit{et al.'s} worm algorithm \cite{boninsegni_prl,boninsegni_pre}) allow one to compute the imaginary-time one-particle Green's function describing the process of (inverse) photoemission. Being the main object studied in many-body perturbation theory~\cite{kadanoff}, the associated one-particle spectral function $A(p,\omega)$ contains the complete electronic excitation spectrum and (local) density of states, which, for dense plasmas, is typically probed using x-ray absorption spectroscopy (XAS), a key diagnostic for condensed matter and materials. XAS has also become a key diagnostic of WDM \cite{Falk2018,Dorchies2011}, to investigate the ionization dynamics \cite{Mercadier2024} as well as the modification of binding energies due to medium effects (ionization potential depression) \cite{Ciricosta2012,Vinko2010}. More recently, resonant inelastic x-ray scattering (RIXS) has been demonstrated as an alternative method for probing the density of states in WDM systems \cite{Forte2024,Sohn2024}.

In the frame of Landau's Fermi-liquid theory, the electronic structure of free (unbound) electrons is often explored in the quasi-particle picture. Generalizing the concept of an effective mass, dynamic many-body effects can be expressed in terms of the self-energy $\Sigma(p,\omega)$, for which approximations are commonly derived using diagrammatic techniques \cite{baym1962}. Besides leading to a finite lifetime of the electron quasi-particle, taking dynamic screening into account, as explored in various flavors of the GW approximation \cite{Holm1998,Tan2018}, introduces satellites to the one-particle spectrum; these may merge into the main peak at finite temperatures when self-consistency of Hedin's equations is enforced~\cite{Fortmann2008}. While being a standard tool for the calculation of excitation spectra in condensed matter theory, GW fails to describe the warm dense UEG beyond weak coupling, giving incorrect results for the total energy \cite{Schoof2015, review}. For strong coupling, better agreement with QMC reference data for the energy can be found using a cumulant expansion \cite{Kas2014,Kas2017,Kas2019}; however, the accuracy of the latter for $A(p,\omega)$ remains unclear.
Previous QMC based investigations of the spectral function of the UEG were restricted to the zero-temperature limit~\cite{Haule2022,Lee2021,Holzmann2023}.


In this Letter, we present non-perturbative results for the single-particle spectral function and the static self-energy for the WDM regime based on extensive \emph{ab initio} PIMC simulations. In addition to being interesting in their own right, our results clearly demonstrate the capability of PIMC to falsify approximate models, conclusively ruling out the existence of pronounced satellite peaks in the UEG at these conditions. Our work opens up the enticing possibility to investigate spectral properties and the density of states of real warm dense matter systems based on highly accurate PIMC simulations, with direct relevance to experiments and theory alike.

\paragraph{\textbf{One-particle Green's function.}} The imaginary-time one-particle Green's function (Matsubara Green's function, MGF), is given in momentum space as
\begin{equation}\label{eq:g_tau}
    G(\vec{p},\tau) = \langle \mathcal{T} \anh(\tau) \cre(0)\rangle,
\end{equation}
and describes the propagation of an electron (or hole) when adding or removing a single electron to/from the system. Both processes are related to the same excitation spectrum, given by the spectral function $A(p,\omega)$,
\begin{equation}\label{eq:inversion}
  G(\vec{p},\tau)= \int\frac{d\omega}{2\pi} e^{-\tau\omega} A(\vec{p},\omega) [ 1 - f(\omega)]\ ,
\end{equation}
where $f(\omega) = 1/[e^{\beta\omega}+ 1]$ is the Fermi function and $\tau\in[0,\beta]$ denotes the \emph{imaginary time} ($\beta=1/k_\textnormal{B}T$). All expectation values are taken in the grand-canonical ensemble where, per convention, all energies are shifted by the chemical potential $\mu$ (which would otherwise explicitly enter in the distribution function).
For the UEG, these quantities only depend on the magnitude $p=|\vec{p}|$.  

\begin{figure}[h]
    \centering
    \includegraphics[width=\linewidth]{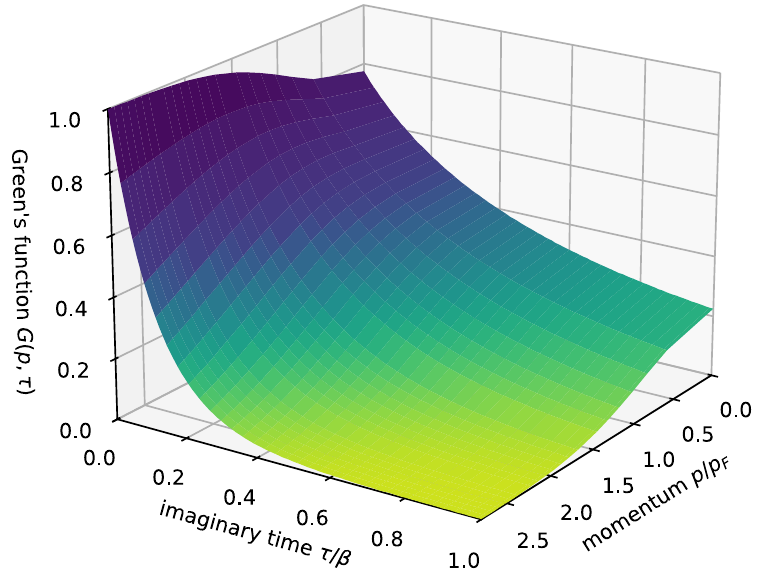}
    \caption{Imaginary-time Green's function at $\theta\approx 2$, $r_s\approx 10$, unpolarized, $\langle N\rangle \approx 20$. Finite-size effects are negligible at these conditions, which agrees with previous investigations of other $p$-resolved properties~\cite{dornheim_prl,Dornheim_PRE_2020}, see also the Supplemental Material~\cite{supplement}.}
    \label{fig:gf_3d}
\end{figure}

The dependence of $G(p,\tau)$ on both arguments is illustrated in Fig.~\ref{fig:gf_3d}, where PIMC results for the unpolarized UEG at $\theta \approx 2 $ and $r_s \approx 10$ are shown. For all values of $p$, the MGF monotonically decays from $\tau=0$ to $\beta$, at which point Eq.~\eqref{eq:g_tau} evaluates to the single-particle density matrix/momentum distribution $n(p)$. Following from the anti-commutation relations, the complementary occupation numbers are obtained at $\tau=0$.

For the non-interacting system, the MGF decays exponentially in between these values, $G(p,\tau) \sim \exp{-\epsilon_p \tau}$. Particles that are added to the system remain in their initial state, the spectral function consists of a delta peak at the single particle energy following a quadratic dispersion $\epsilon_p = p^2/2 - \mu$. This dispersion is modified when taking interaction into account but the ideal behavior still dominates the overall structure of $G(p,\tau)$ at the present conditions. 

\paragraph{\textbf{Self-energy.}}
Being frequency-independent and purely real, the first order self-energy correction amounts to a mere shift of single-particle energies, and is given by the Fock term
\begin{equation}\label{eq:sigma_fock}
\Sigma_\text{Fock}(\vec{p}) = - \int \frac{d\vec{p}'}{(2\pi)^3}  v(\vec{p}-\vec{p}') n(\vec{p}')\ .
\end{equation}
Here, $n(\vec{p})$ is determined self-consistently, as the single-particle energies entering the Fermi function are shifted by $\Sigma$, the chemical potential $\mu$ needs to be adjusted in order to reach a target density.

The negative shift in the dispersion caused by $\Sigma_\text{Fock}$ becomes large for small $p$, which results in an increased occupation of low-momentum states, that has been observed by QMC calculations at high densities \cite{Militzer_PRL_2002,hunger}. When increasing the coupling strength, this phenomenon is counteracted by effects not accounted for on this level of approximation. After reaching a maximum, $n(0)$ eventually drops below the ideal value at strong coupling \cite{dornheim2021momentum}.

For the UEG, Eq.~\eqref{eq:sigma_fock}, given the exact $n(\vec{p})$, already includes all frequency-independent diagrams. Considering higher-order terms, it is useful to split the self-energy into static and correlated part, $\Sigma(p,\omega) = \Sigma_\infty(p) + \Sigma_C(p,\omega)$, with the latter vanishing for $\omega\to \infty$.

Higher order terms do not only lead to a finite lifetime of the quasi-particle, but can also introduce additional features to the spectral function. Characteristic to the UEG is the existence of collective modes (plasmon excitations), which appear when computing the dielectric function $\varepsilon(q,\omega)$ in the random phase approximation (RPA) or beyond. Computing the self-energy on a GW level, i.e., replacing the bare Coulomb term in Eq.~\eqref{eq:sigma_fock} with the dynamically screened interaction $W(q,\omega) = v(q)/\varepsilon(q,\omega)$,
introduces poles to $\Im \Sigma_C$, which are related to the remaining electrons adjusting to the modified charge density when adding/removing an electron. In addition to the quasi-particle peak describing the correlated electron, these poles appear as satellite excitations describing plasmons.

Postponing for now the task of performing the analytic continuation back to real frequency space, it is possible to extract information working directly in the imaginary-time domain. 
Specifically, we consider the frequency moments
\begin{equation}
M_A^{(\alpha)}(p) = \int_{-\infty}^\infty \frac{d\omega}{2\pi}\ \omega^\alpha A(p,\omega)\ ,
\end{equation}
which, due to Eq.~\eqref{eq:inversion} being a two-sided Laplace transform, follow from the derivatives of $G(p,\tau)$ \cite{Dornheim_moments2023}; see the Supplemental Material~\cite{supplement} for additional details.
Following from a high-frequency expansion, the first moment $M_A^{(1)}(p)$ can be related to the frequency-independent (Hartree-Fock) part of the self-energy:
\begin{equation}\label{eq:w1_hf}
    M^{(1)}_A(p) = p^2/2 + \Sigma_\infty(p) + \mu\ ,
\end{equation}
allowing for a straightforward estimation of the exact $\Sigma_\infty(p)$.

\begin{figure}[h]
    \centering
    \includegraphics[width=\linewidth]{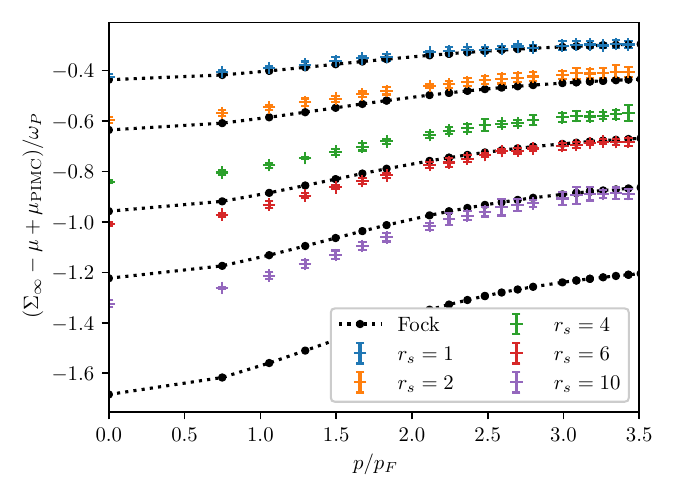}
    \caption{First moment of the spectral function for five densities and the same parameters as in Fig.~\ref{fig:A_rs}. Self-consistent Fock calculations (dotted lines) are compared to PIMC results (colored symbols) that have been obtained directly from the MGF following the procedure from Ref.~\cite{Dornheim_moments2023}, see the Supplemental Material for additional details~\cite{supplement}.}
    \label{fig:ek}
\end{figure}

 In Fig.~\ref{fig:ek}, we compare our new PIMC results for the self-energy to self-consistent Fock calculations. 
Overall, both PIMC and Fock follow the same qualitative trends. As already accounted for by $\Sigma_\text{Fock}$, the negative shift of single-particle energies is most significant at small momenta and vanishes in the limit $p\to \infty$. For $r_s=1$, which corresponds to a strongly compressed state as it is realized, e.g., in spherical implosion experiments at the National Ignition Facility in Livermore~\cite{Tilo_Nature_2023,Dornheim_NatComm_2025}, Fock and PIMC results agree within $2.5\%$. 
With decreasing density, correlation effects increase~\cite{Ott2018}, and the Fock model becomes less accurate; we find deviations of $\sim6\%$ at the metallic density of $r_s=2$ and $~12\%$ at $r_s=4$, which is comparable to the density in experiments with hydrogen jets~\cite{Zastrau}. We thus conclude that the Fock self-energy is not sufficient for the description of many relevant electronic densities, even at the relatively high temperature of $\Theta=2$.
For completeness, we note that Fock completely breaks down for $r_s=10$, which is located at the boundary of the strongly correlated electron liquid, for which we find deviations of $\sim20\%$.


\paragraph{\textbf{Spectral function.}}
As the capstone of our work, we invert Eq.~\eqref{eq:inversion} to reconstruct the one-particle spectral function $A(p,\omega)$ from the quasi-exact PIMC data for $G(p,\tau)$. Being highly sensitive to noise in the input data, this inversion problem is commonly approached using maximum-entropy methods \cite{Gubernatis1991,chuna2025noiselesslimitimprovedpriorlimit} or, when lacking a reasonable default model, by stochastically exploring the space of all possible solutions not invalidated by the PIMC data \cite{Sandvik1998,Mishchenko2001,Vitali2010}. We employ a differential evolutionary algorithm \cite{Nichols2022} as implemented by the \texttt{SmoQyDEAC} code \cite{SmoQyDEAC} to generate a large number of valid solutions for Eq.~\eqref{eq:inversion} and average over them; see the Supplemental Material for additional details~\cite{supplement}.

\begin{figure}[h]
\centering
 \includegraphics[width=\linewidth]{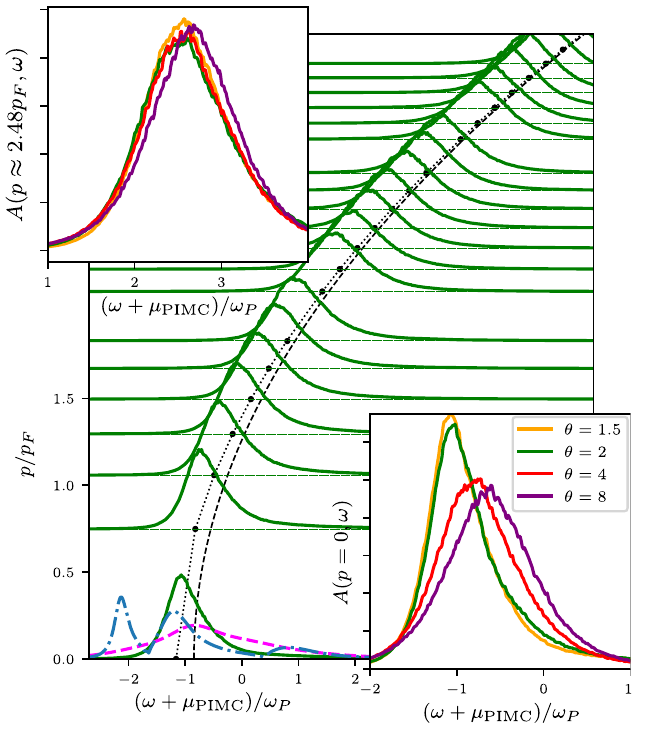}
 \caption{Spectral function at $r_s=4, \langle N\rangle \approx 20$ and $\theta=2$ (unpolarized) for different momenta $p$. The ideal quasi-particle dispersion, $\epsilon_p = p^2/2 - \mu_0$, and one-shot Fock results (direct evaluation of Eq.~\eqref{eq:sigma_fock} using the ideal $n(p)$ and $\mu_0$) are shown by the dotted and dashed black curves, respectively. 
 The dashed magenta and dash-dotted blue curves correspond to the cumulant expansion from Ref.~\cite{Kas2017} and $G_0 W_0$, respectively.
 Insets demonstrate the effect of varying the degeneracy parameter $\theta$, for two momenta.}\label{fig:A_theta}
\end{figure}

Results for the unpolarized UEG at $r_s=4$ are presented in Fig.~\ref{fig:A_theta}. The central part depicts the spectral function for different momenta whereas the insets explore the temperature dependence for two particular values of $p$. At all conditions considered, the spectral function consists of a single peak. Consistent with the discussion of the first moment, its position closely follows the quadratic dispersion of a free particle but experiences a negative shift at small momenta. This effect strongly depends on the temperature, supporting the interpretation as being related to exchange (Fock shows the same trend). At large momenta, position and shape of the quasi-particle peak are only weakly temperature dependent. On the other hand, towards $p=0$, the spectrum is increasingly skewed to the left when lowering the temperature. While its maximum position is shifted, spectral weight located at its shoulders remains largely unaffected as the peak becomes more narrow, leading to an asymmetry. The very same behavior is observed in GW-calculations, in which the quasi-particle peak, shifted by $\Sigma_\text{Fock}$, moves between two plasmon satellite peaks. When enforcing self-consistency, these merge into the main peak and account for the observed shoulders in the spectral shape \cite{Wierling1998}. The same phenomenon is observed using the cumulant expansion considered in Ref.~\cite{Kas2017}, which is shown by the dashed magenta curve for $p=0$ in the main panel of Fig.~\ref{fig:A_theta}.
An explicit comparison between our new PIMC results and previous approximations, including $\mathrm{G}_0\mathrm{W}_0$ and the cumulant expansion, both in frequency and imaginary time domain, is presented in the Supplemental Material~\cite{supplement}, which allows us to conclusively rule out prominent satellite peaks in the spectral function.

\begin{figure}
    \centering
    \includegraphics[width=\linewidth]{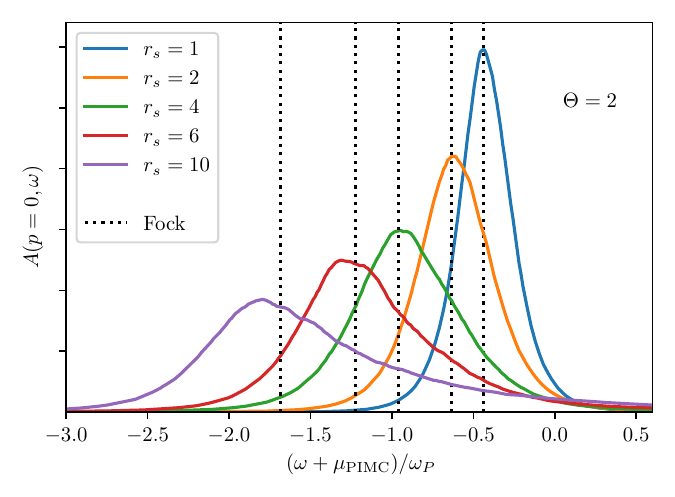}
    \caption{Spectral function of the fully polarized UEG for $p=0$, $\theta=2$ and 5 different coupling parameters $r_s$, $\langle N\rangle \approx 10$. The dotted lines indicate the quasi-particle position predicted by self-consistent Fock calculations.}
    \label{fig:A_rs}
\end{figure}

Let us conclude our investigation by using our new PIMC results to rigorously quantify the impact of correlation effects onto $A(p,\omega)$.
In Fig.~\ref{fig:A_rs}, we show $A(0,\omega)$ covering a broad range of densities.
Overall, we find that spectral weight is progressively shifted towards lower energies when increasing the coupling parameter which is, at least qualitatively, captured by $\Sigma_\text{Fock}$. 
Even at $r_s=10$, the quasi-particle position predicted by Fock is in close vicinity of the maximum, which, due to the growing skewness, however, no longer coincides with the first frequency moment of the spectrum.
We note that all PIMC results for $G(p,\tau)$ and $A(p,\omega)$ are freely available online~\cite{repo} and can be used in future works to assess the accuracy of existing models and approximations and to guide the development of new methodologies.

\paragraph{\textbf{Summary and Outlook.}}
We have carried out quasi-exact PIMC simulations of the UEG over a broad range of densities, reaching from weak ($r_s=1$) to strong coupling ($r_s=10$) to compute the Matsubara Green's function $G(p,\tau)$. This has allowed us to present, to our knowledge, the first approximation-free results for the the self-energy shift $\Sigma_\infty(p)$ and the one-particle spectral function $A(p,\omega)$.
While the PIMC results show similar trends w.r.t.~the dependence on momentum, coupling strength and temperature -- self-consistent Fock calculations seem to give an almost perfect prediction for the peak position -- $\mathrm{G}_0 \mathrm{W}_0$ greatly overestimates the damping, even at moderate densities. At the temperatures considered, no distinct satellite structure is observed.

We are convinced that our results will provide useful benchmarks for the development of analytical models and more advanced self-energy approximations, including improved cumulant models \cite{Kas2017}. Given the rather simple structure of the resulting spectra, this may be facilitated by going beyond Eq.~\eqref{eq:w1_hf} and to extract dynamic self-energies from the PIMC data, possibly resulting in a parameterization.

Most importantly, the present approach is not restricted to the UEG. 
Indeed, full two-component PIMC simulations
of real WDM systems such as hydrogen~\cite{Filinov_PRE_2023,Dornheim_MRE_2024} and beryllium~\cite{Dornheim_NatComm_2025} have been demonstrated very recently, and corresponding results for $G(p,\tau)$ and $A(p,\omega)$ will allow one to directly investigate the influence of many-body effects on the band gap, ionization potential depression and ionization state~\cite{Bellenbaum_PRR_2025,Bonitz_CPP_2025,Hu2017} with immediate consequences for the interpretation of experiments. 
Finally, computing the density of states for fixed ionic positions (snapshot) might potentially provide a particularly rigorous benchmark for the improvement of density functional theory calculations, which yield incorrect results for band gaps when dynamic screening is taken into account using a static exchange-correlation kernel~\cite{Onida2002}.

\section*{Acknowledgements}

\begin{acknowledgements}
\noindent
P.H. gratefully acknowledges useful discussions with Pontus Svensson.

This work has received funding from the European Research Council (ERC) under the European Union’s Horizon 2022 research and innovation programme (Grant agreement No. 101076233, "PREXTREME"). 
Views and opinions expressed are however those of the authors only and do not necessarily reflect those of the European Union or the European Research Council Executive Agency. Neither the European Union nor the granting authority can be held responsible for them.
We gratefully acknowledge funding from the Deutsche Forschungsgemeinschaft (DFG) via projects DO 2670/1-1 and BO1366/13-2.

Computations were performed at the Norddeutscher Verbund f\"ur Hoch- und H\"ochstleistungsrechnen (HLRN) under grant mvp00024.
\end{acknowledgements}

\bibliography{bibliography}

\clearpage
\FloatBarrier
\onecolumngrid
\section{Supplementary Material}

\subsection{PIMC computation of the imaginary-time Green's function}
The configurational space sampled by the worm algorithm \cite{boninsegni_prl,boninsegni_pre} is composed of two parts, $Z_W = Z + Z'$.
The closed trajectory sector $Z$ corresponds to the grand-canonical partition function:
$$ Z = \sum\limits_{N=0}^\infty e^{\beta\mu N} Z_N \,, $$
where the $N$-particle sector $Z_N$ is given by the sum over all (anti-)symmetrized $N$-particle trajectories, returning to the same position, $\vec{R}_P = \vec{R}_0$, after traversal of the imaginary time interval $\tau \in [0,\beta]$:
$$ Z_N = {\frac{1}{N!} \sum\limits_{\pi} (\pm 1)^{\pi_l}} \int d\vec{R}_0 \ldots d\vec{R}_{P-1} \prod\limits_{i=0}^{P-1} \rho(\vec{R}_i,{\hat{\pi}}\vec{R}_{i+1};\epsilon) \,.$$
The open sector $Z'$ additionally includes a single trajectory (``worm'') with open ends (``head'' and ``tail'' of the worm), with a weight given by the coordinate space one-particle Green's function:
$$ Z' = Z C \sum_{ij} \int d\vec{r}_{\Psi^\dagger} d\vec{r}_{\Psi}  G(\vec{r}_{\Psi^\dagger},\vec{r}_{\Psi},\tau_i-\tau_j) \, .$$
Here $C$ is a free parameter that can be adjusted to control how much time the algorithm spends in each sector. Considering a homogeneous system in equilibrium, $G$ only depends on the relative time argument and spatial distance between head and tail. Averaging over the position of the tail in space and time and switching to momentum space, one arrives at the Green's function estimator given in Ref.~\cite{boninsegni_pre}:
$$ G(\vec{p},\epsilon j) = \frac{ \langle \delta^{Z'} \rangle_{Z_W} }{\langle \delta^Z \rangle_{Z_W}} \frac{\langle \delta_{j,(j_\text{head}-j_\text{tail})} e^{i\vec{p}(\vec{r}_\text{head}-\vec{r}_\text{tail})}\rangle_{Z'}}{CVP}\, .$$
The prefactor, determining the ratio between the number of encountered closed $\langle \delta^Z\rangle_{Z_W}$ and open $\langle \delta^{Z'}\rangle_{Z_W}$ configurations, corresponds to the ratio of partition functions $Z'/Z$ and introduces the correct normalization, as the Green's function is defined as a grand-canonical expectation value, normalized to $Z$.
\begin{figure}[h]
    \centering
    \includegraphics[width=0.45\linewidth]{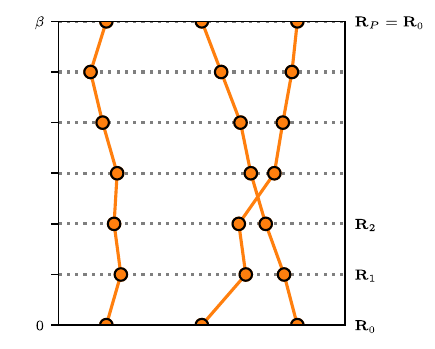}
    \includegraphics[width=0.45\textwidth]{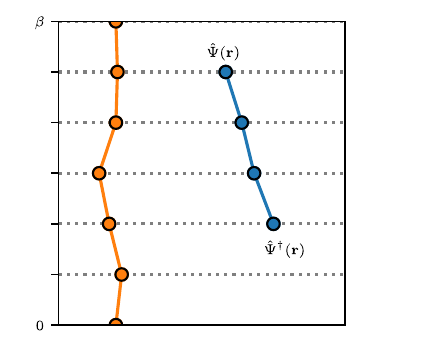}
    \caption{PIMC configurations involving only closed trajectories (left panel) and a worm (right panel).}
    \label{fig:worm_configuration}
\end{figure}
Fig.~\ref{fig:worm_configuration} illustrates the two classes of configurations sampled by the worm algorithm. The left panel shows a closed configuration consisting of three particles, two of which are involved in an exchange cycle. The three particles starting at $\vec{R}_0$ at $\tau=0$, propagate through in the imaginary time, until they reach their initial position $\vec{R}_P = \vec{R}_0$ at $\tau=\beta$. The configuration shown in the right panel on the other hand contains a worm. In addition to the closed trajectory shown in orange, a new particle (blue) enters the system at one point in the imaginary time, before eventually being removed again.

\subsection{Spectral function in the $G_0W_0$ approximation}

An accurate description of the UEG has to take screening into account. The mutual influence of two test charges does not follow the bare Coulomb potential, and is affected by all other electrons dynamically adjusting to the modified charge density. A prominent many-body approximation that takes dynamical screening into account is the GW approximation that follows by replacing the static Coulomb potential in Eq.~\eqref{eq:sigma_fock} with the dynamically screened potential, $W(1,2)=V(1,2) \varepsilon^{-1}(1,2)$:
\begin{equation}\label{eq:sigma_gw}
\Sigma_\text{GW}(1,1') = i G(1,1')  W(1,1'),
\end{equation}
where $1 \equiv (\vec{r}_1,t_1)$ denotes a set of time and space coordinates.
In his seminal work \cite{hedin}, Hedin proposes to determine $W$ and $G$ self-consistently, i.e. consider $W$ as a functional of $G$. In lowest order, this amounts to computing the irreducible susceptibility in the random phase approximation:
\begin{equation}\label{eq:pigg}
 \Pi(1,2) \approx  -i G(1,2) G(2,1),
\end{equation}
from which then $W$ entering Eq.~\eqref{eq:sigma_gw} follows using $\varepsilon(\vec{q},\omega)=1-v_\vec{q} \Pi(\vec{q},\omega)$.
This approximation is expected to be valid at high densities, where due to strong screening a particle and a hole are expected to propagate independently of each other. On the other hand, for dilute systems (strong coupling), their interaction becomes more important than the effect of the medium, which is taken into account e.g. in the ladder approximation \cite{Kremp1984}.

Fully self-consistent calculations for the UEG at zero temperature have been first explored by Holm and von Barth \cite{Holm1998}. While these show a better agreement with the total energy as determined from exact Monte Carlo simulations, the polarization function following from the self-consistent solution for $G$ leads to unphysical results for the dielectric function, in particular it violates the f-sum rule. This problem can be evaded by keeping $W$ as computed from $G_0$ fixed and restricting the self-consistency cycle to the calculation of $\Sigma$ and $G$, a procedure referred to as $\mathrm{GW}_0$. Alternatively, the idea of self-consistency may be abandoned altogether by directly evaluating  Eq.~\eqref{eq:sigma_gw} using the ideal Green's function, which is referred to as $\mathrm{G}_0\mathrm{W}_0$. For the UEG at finite temperature, (partially) self-consistent calculations have been carried out in Ref.~\cite{Fortmann2008}.

To get an impression of the effect dynamical screening has on the spectral function, we compute the self-energy in the $\mathrm{G}_0\mathrm{W}_0$ approximation. In momentum and frequency space Eq.~\eqref{eq:sigma_gw} is a convolution:
\begin{equation}\Sigma_\text{GW}^\gtrless(\vec{p},\omega)=i\int\limits\frac{d\vec{ q}}{(2\pi)^3}\int\limits_{-\infty}^{\infty}\frac{d\omega'}{2\pi} G^\gtrless(\vec{p}-\vec{q},\omega-\omega') W^\gtrless(\vec{q},\omega').\end{equation}
Following Ref.~\cite{kremp} we eventually arrive at a retarded self-energy of:
\begin{equation} \operatorname{Im}\Sigma(\vec{p},\omega)=\int\frac{d\vec{q}}{(2\pi)^3}\int\limits_{-\infty}^\infty\frac{d\omega'}{2\pi}  \operatorname{Im}\frac{v_\vec{q}}{\varepsilon(\vec{q},\omega')} A(\vec{p}-\vec{q},\omega-\omega') \{1+f_B(\omega')-f_F(\omega-\omega')\}.
\end{equation}
This expression further simplifies when inserting the ideal spectral function, allowing us to switch to spherical coordinates and explicitly carry out the angular part of the integration:
\begin{equation}\label{eq:gw}
\operatorname{Im}\Sigma(p,\omega)=\frac{1}{\pi p}\int\frac{dq}{q}\int\limits_{\omega-(p+q)^2/2}^{\omega-(p-q)^2/2}d\omega' \operatorname{Im}\frac{1}{\varepsilon(q,\omega')}  \{1+f_B(\omega')-f_F(\omega-\omega')\} \,.
\end{equation}
The real part of the self-energy follows from the Kramers-Kronig relations:
\begin{equation}\label{eq:kk_sigma}
\operatorname{Re}\Sigma(p,\omega)=\operatorname{Re}\Sigma(p,\infty)-P\int\frac{d\omega'}{\pi}\frac{\operatorname{Im}\Sigma(p,\omega')}{\omega-\omega'} \, ,
\end{equation}
where $\operatorname{Re}\Sigma(p,\infty)$ is the constant (frequency-independent) part given by the Fock term.

\begin{figure}
    \centering
    \includegraphics[width=0.8\linewidth]{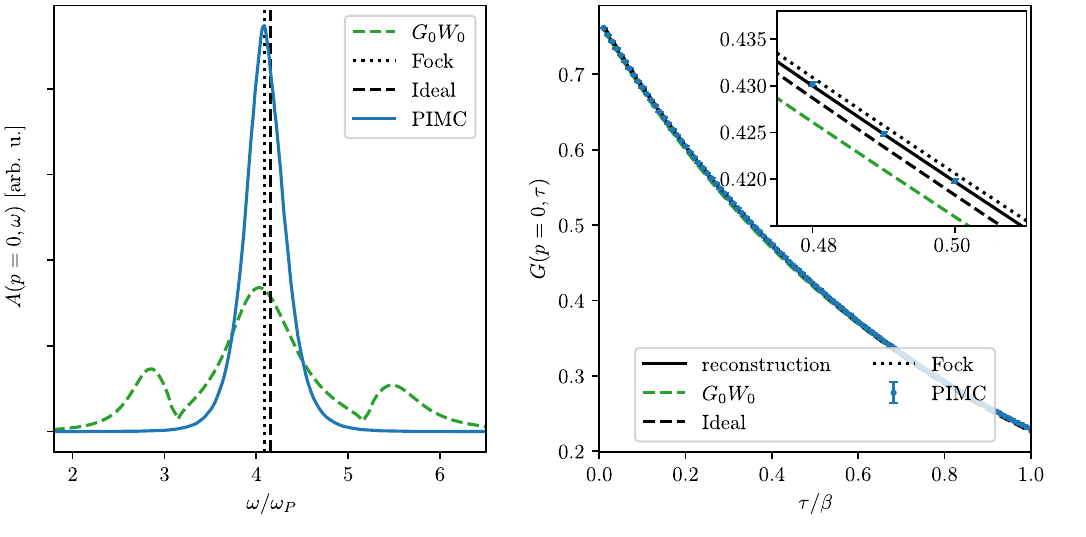}
    \includegraphics[width=0.8\linewidth]{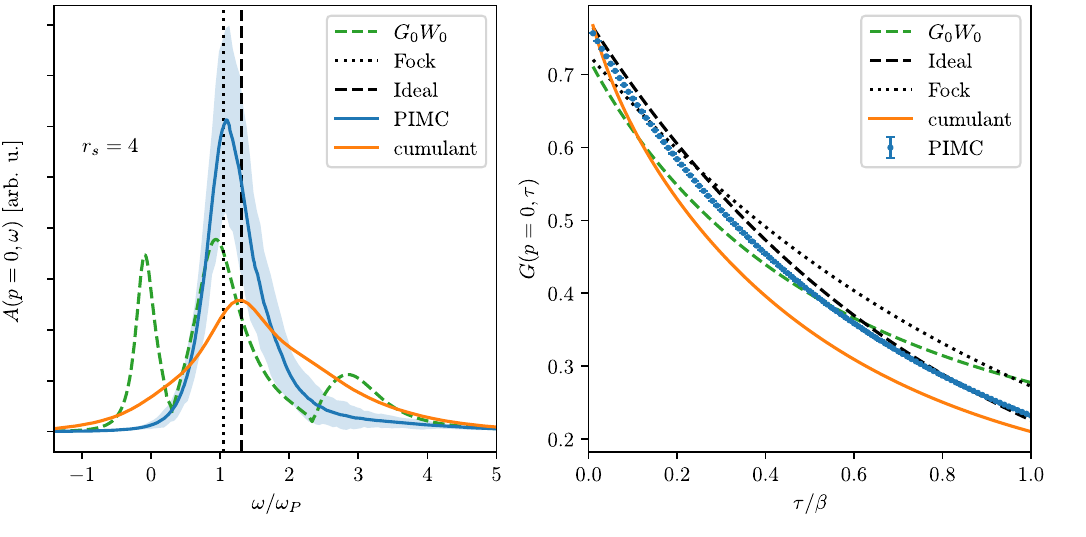}
    \caption{Spectral function and imaginary time Green's function for the polarized UEG at $r_s=1$ and $\theta=2$ (top) and unpolarized UEG at $r_s=4$ and $\theta=2$ (bottom). PIMC results (blue) are compared to the ideal (black, dotted), Fock (black, dashed) and $G_0W_0$ (green, dashed) result. The inset in the upper right panel additionally shows the MGF computed from the reconstructed spectral function (black, solid). For $r_s=4$ we additionally show results obtained using the cumulant expansion from Ref.~\cite{Kas2017} (orange).}
    \label{fig:pimc_gw}
\end{figure}

Results for the spectral function as well as the imaginary-time Green's function are shown in Fig.~\ref{fig:pimc_gw}. As before, we compare the PIMC data (and reconstruction result) to self-consistent Fock calculations (dotted), the result for the free electron gas is shown by the dashed lines. Results for the spectral function computed from the $\mathrm{G}_0\mathrm{W}_0$ self-energy are shown by the green dashed line.
It is important to note that computing the density from the GW result will lead to a different value, since in one-shot calculations, the chemical potential of the free electron gas needs to be used in the Lindhard function to obtain the correct plasmon position at $p=0$.
The upper panels show results for the polarized UEG at $r_s=1$ and $\theta=2$, both in the frequency (left) and imaginary-time domain (right). At the present conditions, interaction effects are barely noticeable in the Green's function, which for the most part follows the exponential decay with the kinetic energy. Variations between the different approaches become only visible in the inset. As already discussed before, the Fock correction amounts to a minuscule shift of spectral weight towards lower energies, a phenomenon which is also observed in the PIMC data and leading to very good agreement at small and large $\tau$. However, larger differences are observed at intermediate $\tau$. Due to the finite width of the peak,  $G_\text{PIMC}(p,\tau)$ is given by a superposition of exponential functions decaying with a slightly different rate.
This effect is greatly amplified when considering the spectral function computed in $G_0 W_0$. As significant portions of spectral weight are shifted to energies above and below the quasiparticle peak, we find a more complex behavior in the imaginary time that can be roughly grouped into three modes showing exponential decay with different rates. This results in larger discrepancies towards the PIMC data.
At small values of $\tau$, the GW result for $G(p,\tau)$ lies below all other curves and deviates in the other direction for large $\tau$. This is not surprising as even at densities as high as $r_s=1$, GW gives increasingly inaccurate predictions for the interaction energy as well \cite{review}. We conclude that $G_0W_0$ gives an inadequate description of the spectral function at all parameters considered. Deviations become even more significant when turning to lower densities/stronger coupling ($r_s=4$, bottom panels), as more spectral weight moves to the satellite peaks. Additionally shown are results from Ref.~\cite{Kas2017} (orange), which have been obtained using a cumulant expansion. These -- at least qualitatively -- give a better agreement with the spectral shape observed in the PIMC results, but overestimate the damping as well. In the imaginary-time domain, the cumulant results match the PIMC data better than $\mathrm{G}_0\mathrm{W}_0$ at $\tau=0$ and $\tau=\beta$. However, they still show a similar level of disagreement as $\mathrm{G}_0\mathrm{W}_0$ when considering the whole $\tau$-range. As expected, the MGF related to the reconstructed spectral function, shown in the upper right panel by the solid black line, perfectly matches the PIMC data within the error bars.

\subsection{Finite size effects}

\begin{figure}[h]
    \centering
    \includegraphics[width=0.48\linewidth]{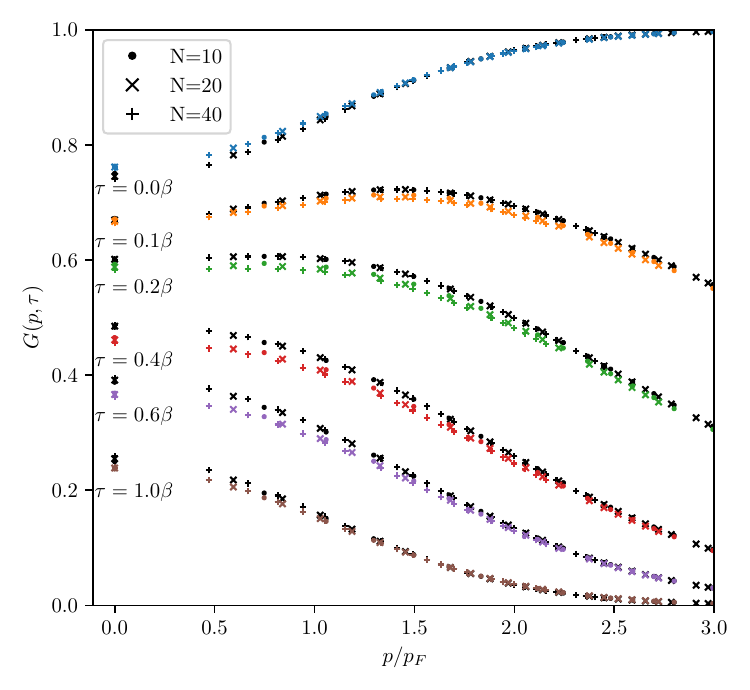}
    \includegraphics[width=0.48\linewidth]{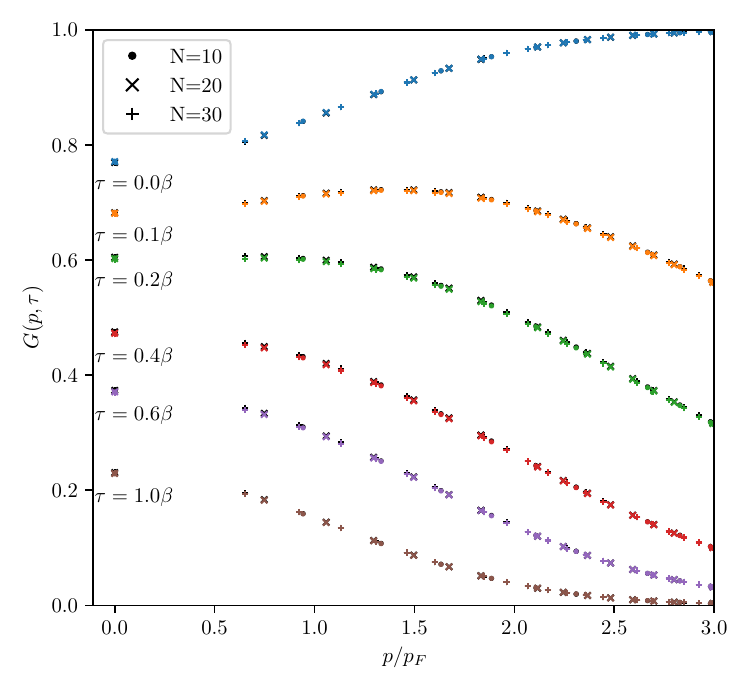}
    \caption{Matsubara Green\textcolor{blue}{'s} function $G(p,\tau)$ for selected imaginary-time arguments for different average numbers of particles $N$. (Left: $r_s=6$, spin-polarized. Right: $r_s=1$, unpolarized. Both $\theta=2$). Colored: PIMC results, Black: Fock}
    \label{fig:finite_size}
\end{figure}

All PIMC simulations that have been presented in this work have been performed in the grandcanonical ensemble, meaning that the volume of the cubic simulation cell $\Omega=L^3$, inverse temperature $\beta$ and chemical $\mu$ are being fixed, uniquely determining the mean particle number $\langle N\rangle$ present in the box. In practice, the main obstacle to simulations of large systems is the fermion sign problem \cite{dornheim_sign_problem,Dornheim2021}, leading to the computational effort required growing exponentially with $\langle N\rangle$ at any given average density and temperature. Considering the UEG, the repulsive Coulomb interaction leads to electrons being more strongly localized with increasing coupling parameter $r_s$. At a given value of the degeneracy parameter $\theta$, the sign problem becomes most severe in the non-interacting (high-density) limit $r_s \to 0$, where consequently the feasible system size is the smallest. However, ultimately we are interested in the behavior of the UEG in the thermodynamic limit, where we simultaneously take the limits of $N\to\infty$ and $\Omega\to\infty$ with the density $N/\Omega$ being kept constant.
Differences between PIMC simulations for finite $N$ and the thermodynamic limit are often denoted as \emph{finite-size effects} and have been studied extensively in the literature, in particular for the UEG~\cite{Chiesa_PRL_2006,Drummond_PRB_2008,Holzmann_PRB_2016,dornheim_prl,Dornheim_PRE_2020,Dornheim_JCP_2021}.
In particular, finite-size corrections for integrated properties such as the interaction energy often make use of the fact that wave-number resolved quantities such as the static structure factor and density response do not exhibit any significant dependence on the system size except for the obvious momentum quantization in the finite simulation box. This phenomenon extends to imaginary-time quantities, as investigated for the two-particle Green's function in Ref.~\cite{Dornheim_PRE_2020}. In order to obtain the same finite-size scaling when computing the interaction energy from the one-particle spectral function instead of the static structure factor, i.e. carrying out a momentum and frequency space integral over $A(p,\omega)$ given by the Galitskii-Migdal formula \cite{kadanoff}, the same needs to hold for momentum-resolved one-particle properties. 

In Fig.~\ref{fig:finite_size}, we confirm this behavior for the Matsubara Green\textcolor{blue}{'s} function at two representative sets of conditions, namely $r_s=6$ (left) and $r_s=1$ (right) at $\Theta=2$.
Specifically, we show the dependence on the wave number $p$ for different (fixed) imaginary-time arguments $\tau$, with the dots, crosses and pluses showing results for different average numbers of particles $N$. The colored and black symbols show PIMC and Fock results.
For both data sets, the main effect of the system size is indeed the $p$-grid, whereas any intrinsic finite-size effects are negligible.

\subsection{Computational details}
All PIMC results have been computed using the implementation of the worm algorithm provided by the open-source \texttt{ISHTAR} code~\cite{ISHTAR}. In order to compute fermionic observables we rely on the direct PIMC method, i.e. configurations are sampled according to their bosonic weights and fermionic results are computed via reweighting. While this avoids introducing a systematic bias with regards to the choice of the trial nodal surface required for restricted PIMC simulations, the range of applicability is ultimately limited by the fermion sign problem (FSP), which leads to a growing cancellation of positive and negative contributions as exchange becomes important, i.e. at high densities and low temperatures \cite{troyer,dornheim_sign_problem}. Simulations become particularly inefficient when additionally allowing the particle number to fluctuate as required to measure grand-canonical expectation values, as the overlap between bosonic and fermionic particle number distribution progressively vanishes at strong degeneracy or when increasing the particle number~\cite{dornheim_sign_problem,Hamann2025}. To give a specific example, the mean values of the sign encountered in the simulations corresponding to the right panel of Fig.~\ref{fig:finite_size} ($r_s=1$, $\theta=2$) are $\langle s\rangle \approx 0.45$ for $\langle N\rangle=10$, $\langle s\rangle \approx 0.21$ for $\langle N\rangle=20$ and $\langle s\rangle \approx 0.10$ for $\langle N\rangle=30$. Since even at a moderate sign immense computational resources are required to determine $G(p,\tau)$ with the necessary accuracy required to attempt the spectral reconstruction -- a typical calculation with e.g. $\langle s\rangle \approx 0.5$ for $\langle N\rangle = 20$ at $\theta=2$, $r_s=4$, see Fig.~\ref{fig:A_theta}, consumes about  $1\,\mathrm{MCPUh}$ -- we restrict ourselves to $\Theta \gtrsim 1.5$ in this work.

\begin{figure}[h]
    \centering
    \includegraphics[width=0.96\linewidth]{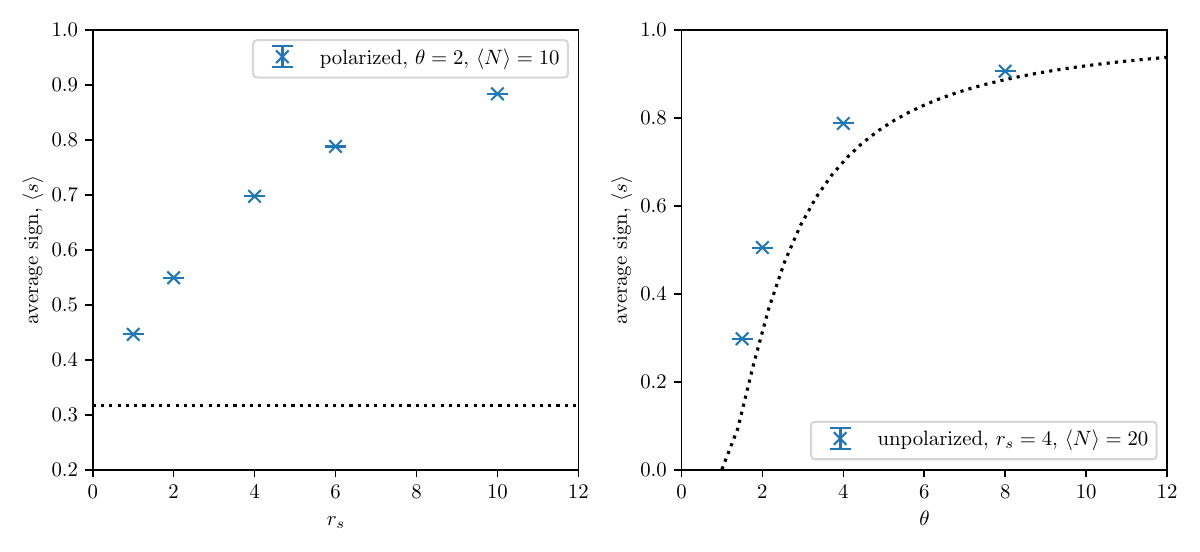}
    \caption{Average sign occurring in the PIMC simulations corresponding to Fig.~\ref{fig:A_rs} (left) and Fig.~\ref{fig:A_theta} (right). The average sign for the non-interacting system at the same parameters is shown by the dotted line.}
    \label{fig:sign}
\end{figure}

Simulation results for the average sign are shown in Fig.~\ref{fig:sign}. We can see in the left panel, which shows results for a fixed degeneracy parameter $\theta=2$ and average particle number, $\langle N\rangle = 10$), how the PIMC result transitions between the non-interacting limit at $r_s \to 0$ (ideal Fermi gas, shown by the dotted line) and classical value of $\langle s\rangle = 1$ as the system becomes more strongly coupled, leading to a reduction of permutation cycles occurring in the simulation. The right panel on the other hand shows results for different values of the degeneracy parameter $\theta$ with $r_s=4$ being kept constant. Again, compared to the ideal Fermi gas (dotted), the sign problem of the interacting system is overall less severe. In the high-temperature limit the average sign asymptotically approaches unity. With decreasing $\theta$ the sign drops rapidly and (for the non-interacting system) reaches zero at $\theta\approx 0.989$. Whereas canonical simulations of a sufficiently small number of fermions are straightforwardly possible for $\theta < 1$, the grand-canonical bosonic partition function (and particle number) diverges for $\mu \geq 0$, making it impossible to compute fermionic observables via reweighing.

Having PIMC data for $G(p,\tau)$ at a broad range of parameters at hand, the imaginary-time data and associated uncertainty levels can then be used as input in various reconstruction schemes to invert Eq.~\eqref{eq:inversion}. As a strategy, we decide to use a stochastic method to optimize a large number of trial solutions for $A(p,\omega)$ until they reproduce the PIMC data for $G(p,\tau)$ at all known values of $\tau$ within the error bars, at which point valid solutions are stored to prevent overfitting. Averaging over the results allows to suppress noise originating from the numerical instability of the inversion problem without the need of having to introduce a default model or any other form of regularization, possibly biasing a spectrum of apriori unknown shape. A highly efficient procedure for sampling the space of all possible solutions is given by the Differential Evolution for Analytic Continuation (DEAC) method introduced in Ref.~\cite{Nichols2022}. We use the open-source implementation provided by the \texttt{SmoQyDEAC} Julia package \cite{SmoQyDEAC}. As demonstrated in Fig.~\ref{fig:pimc_gw}, the MGF computed from the resulting average spectrum is in perfect agreement with the PIMC data.

\subsection{Extraction of frequency moments}
Being related to the Green's function via a two-sided Laplace transform, frequency moments (Eq.~\eqref{eq:A_moments}) of $A(p,\omega)$ follow from the derivatives of $G(p,\tau)$ (see Ref.~\cite{Dornheim_moments2023} where the equivalent relationship for the dynamic structure factor is explored in great detail):
\begin{eqnarray}\label{eq:A_moments}
M^{(\alpha)}_A(p) = \left(-1\right)^\alpha 
 \left\{\left.
\frac{\partial^\alpha }{\partial\tau^\alpha} G(p,\tau)
\right|_{\tau=0}\left. +
\frac{\partial^\alpha}{\partial\tau^\alpha}
G(p,\tau)
\right|_{\tau=\beta}
\right\}.
\end{eqnarray}
To numerically evaluate the derivatives of the PIMC data for $G(p,\tau)$, which, due to their stochastic origin, are affected by noise,
we fit a trial function to a window of data points around $\tau = 0$, respectively $\tau = \beta$. While in theory any functional form which can be put into agreement with the data is suitable, in practice one has to balance the number of free parameters and window length to avoid both under- and overfitting.
\begin{figure}[h]
    \centering
    \includegraphics[width=0.49\linewidth]{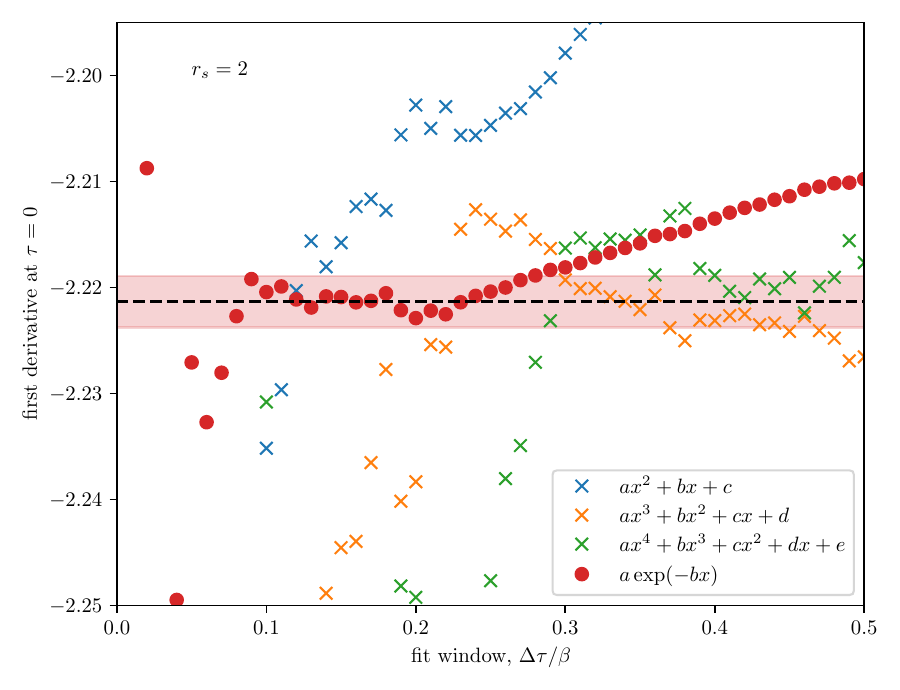}
    \includegraphics[width=0.49\linewidth]{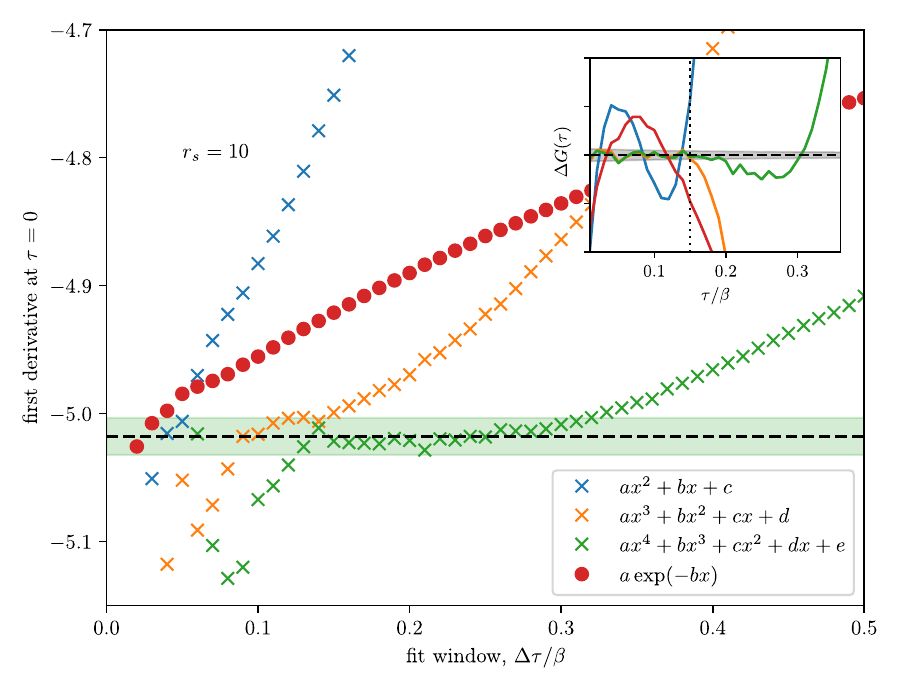}
    \caption{First derivative of the MGF at $\tau=0$ determined by fitting different trial functions to a varying range of PIMC data points. The inset in the right panel shows the deviation from the PIMC data for a window size of $\Delta \tau = 0.15\beta$.}
    \label{fig:moments_fit}
\end{figure}
Fig.~\ref{fig:moments_fit} demonstrates how the obtained value for the derivative depends on the choice of window length, for both an exponential function and polynomial trial functions of varying degree. Let's start by discussing the left panel, which shows results for $\theta=2$, $r_s=2$ and $p=0$. The results obtained using the exponential trial function (red circles) can be roughly categorized into three regions. Up to a window length of $\Delta \tau \approx 0.1\beta$ one is overfitting the statistical noise of the PIMC data. Slight adjustments of the fitting window lead to fluctuating results. For windows larger than $\Delta \tau \approx 0.2\beta$, the MGF can no longer be accurately approximated by a single exponential function, one is underfitting the PIMC data. The estimated derivative increases monotonously with every additional data point taken into account. In between these two regions, from about $\Delta\tau = 0.1\beta$ to $\Delta \tau = 0.2$, the estimated derivative reaches a stable plateau, where the trial function fits the PIMC data well. The remaining fluctuation in this region gives us an estimate for the uncertainty of the fit, which is shown by the shaded area (three $\sigma$).\\ Turning to polynomial fits, we can see that with each additional term included a larger fitting window needs to be used for the first derivative to become stable (overfitting occurs up to a higher $\Delta \tau$). At the same time a polynomial of high order is required to capture the (at weak coupling) dominantly exponential decay over large intervals (prevent underfitting the data). At the present conditions, no stable region of window sizes can be identified for quadratic fits. The first derivative estimated using a cubic fit increases while fluctuating up to about $0.2\beta$, then behaves more stable and drifts in a range of about $6\sigma$ around the value estimated using the exponential fit. The same is true when using a quartic fit (green crosses), which converges at about $\Delta \tau = 0.3\beta$. When further increasing the fit window to larger values than shown in the figure, both cubic and quartic fits start to underfit the data as well, leading to monotonously increasing estimates for the first derivative. As stable fitting window sizes can be identified most clearly using an exponential trial function, we consider it to be an appropriate choice to compute the derivative. The situation changes drastically when turning to stronger coupling (right panel, $r_s=10$), where the MGF even on intermediate time scales can no longer be modeled as a single exponential function. Despite starting from a value that lies close to what is taken to be the final estimate, both exponential and quadratic fit immediately underfit the data and yield results that monotonously grow with the window size. Only when using a quartic or higher polynomial we can identify a plateau at which the estimated derivative becomes independent of the window size. Results shown in Fig.~\ref{fig:ek} were obtained using an exponential trial function for $r_s \in \{1,2,4\}$ and a fifth-order polynomial for $r_s \in \{6, 10\}$. The suitable range of window sizes was chosen by hand for each $p$-point.

\subsection{Testing the existence of a satellite structure}
\begin{figure}[h]
    \centering
    \includegraphics[width=\linewidth]{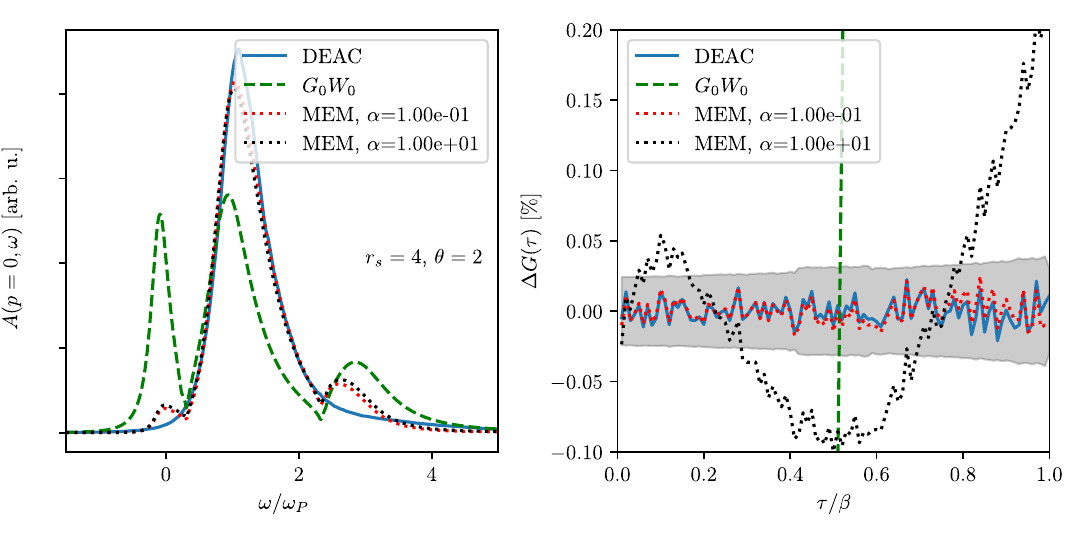}
    \includegraphics[width=\linewidth]{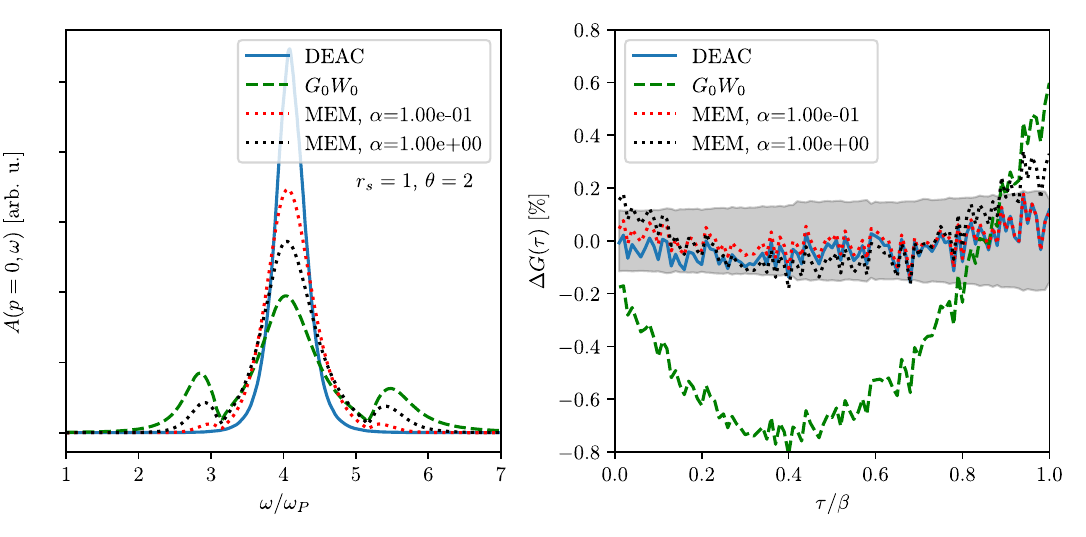}
    \caption{Reconstructed spectra and deviation in the imaginary-time domain when using the maximum entropy method (MEM) taking $G_0 W_0$ as the default model. Shown are results for two different choices of the regularization parameter $\alpha$. Top: $r_s=4$, $\theta=2$, $N=20$ (unpolarized), Bottom: $r_s=1$, $\theta=2$, $N=10$ (polarized).}
    \label{fig:mem}
\end{figure}

To investigate whether a satellite structure can be put into agreement with the imaginary-time PIMC data, we apply the Maximum entropy method \cite{Chuna2025} taking the $G_0 W_0$ result as a Bayesian prior $D(\omega)$ to carry out the analytic continuation, which means finding the spectrum $A_\text{trial}(\omega)$, that solves the least-square problem
$$ \min \frac{1}{2} \sum_i \left[ G_\text{trial}(\tau_i)  - G_\text{PIMC}(\tau_i) \right]^2 + \alpha S[ A_\text{trial} | D ], $$
with $S[A_\text{trial}| D]$ being the Shannon-Janes entropy, weighted by the regularization parameter $\alpha$, that controls how much the default model, i.e. the three-peaked structure predicted by $G_0W_0$ should be favored. This parameter needs to be chosen carefully. If too large, one underfits the PIMC data, eventually recovering the default model. If $\alpha$ is chosen too low, MEM instead overfits the PIMC data and yields spectra with unphysical noise. To check whether the PIMC data falsifies the existence of satellite peaks, we want to keep $\alpha$ as large as possible, but low enough, that the resulting MGF stays within the uncertainty margins of the PIMC data. Resulting spectra and relative deviation from the PIMC results in the imaginary-time domain are shown in Fig.~\ref{fig:mem} for two different densities and two values of $\alpha$ each. Let's start by considering the upper panels. In the frequency domain (left) there are only minor differences between the spectra obtained using $\alpha=0.1$ and $\alpha=10$. Compared to the $G_0 W_0$ solution, almost all spectral weight is removed from the lower plasmon satellite and transferred to the quasi-particle peak, leaving only a minor bump behind. Spectral weight still persists in the range of the upper satellite (the same is true for the DEAC solution). Compared to the default model, the upper satellite however starts to wash out and merges with the central peak. Turning to the right panel which shows the relative deviation from the PIMC data, we can see that as discussed before, the $G_0 W_0$ result significantly violates the PIMC data. The MEM result for $\alpha=10$ gives a better agreement, but still overshoots the uncertainty margins by up to a factor of about four. Only the DEAC and MEM result for $\alpha=0.1$ reproduce the PIMC data within the uncertainty margins. The situation is even more clear when turning to weaker coupling ($r_s=1$) as shown in the bottom panels. With decreasing values of $\alpha$, the MEM spectra continually loose spectral weight in the satellites and approach the DEAC result. We conclude that the imaginary-time data does not indicate the presence of satellite peak.

\end{document}